\documentclass[11pt]{article}
\usepackage{epsfig}        

\begin{document}
\newcommand{\writingdate}{July 5th, 2000}
\newcommand{\I}{{\rm I}}
\newcommand{\II}{{\rm I\!I}}
\newcommand{\III}{{\rm I\!I\!I}}
\newcommand{\room}{\rule[-0.3cm]{0cm}{0.8cm}}
\def\no{{\noindent}}
\def\p{\mathop \partial}
\newcommand{\hsp}{\hspace*{3mm}}
\newcommand{\vsp}{\vspace*{3mm}}
\newcommand{\be}{\begin{equation}}
\newcommand{\ee}{\end{equation}}
\newcommand{\bd}{\begin{displaymath}}
\newcommand{\ed}{\end{displaymath}}
\newcommand{\sgn}{{\rm sgn}}
\newcommand{\IM}{{\rm Im}}
\newcommand{\RE}{{\rm Re}}
\newcommand{\bra}{\langle}
\newcommand{\ket}{\rangle}
\newcommand{\bigbra}{\left\langle}
\newcommand{\bigket}{\right\rangle}
\newcommand{\order}{{\cal O}}
\newcommand{\minus}{\!-\!}
\newcommand{\plus}{\!+\!}
\newcommand{\inn}{\!\cdot\!}
\newcommand{\hx}{\hat{x}}
\newcommand{\hy}{\hat{y}}
\newcommand{\hz}{\hat{z}}
\newcommand{\bv}{\mbox{\boldmath $v$}}
\newcommand{\bOmega}{\mbox{\boldmath $\Omega $}}
\newcommand{\bxi}{{\mbox{\boldmath $\xi$}}}
\newcommand{\beps}{\Delta\mbox{\boldmath $J$}}
\newcommand{\bk}{\mbox{\boldmath $k$}}
\newcommand{\ba}{\mbox{\boldmath $A$}}
\newcommand{\bw}{\mbox{\boldmath $w$}}
\newcommand{\bPhi}{\mbox{\boldmath $\Phi$}}
\newcommand{\bJ}{\mbox{\boldmath $J$}}
\newcommand{\bB}{\mbox{\boldmath $B$}}
\newcommand{\bBS}{\mbox{{\boldmath $B$}}^*}
\title{\bf
Non-Deterministic Learning Dynamics in
Large Neural Networks due to Structural Data Bias}
\author{\bf H.C. Rae $~~~~$ J.A.F. Heimel $~~~~$ A.C.C.
Coolen\\[3mm]
Department of Mathematics, King's College London \\ The Strand, London WC2R 2LS,
UK}
\date{\writingdate}
\maketitle
\begin{abstract}
\noindent
We study the dynamics of on-line learning in large
$(N\to \infty )$ perceptrons, for the case of training
sets with a structural $\order(N^0)$ bias of the input vectors, by
deriving exact and closed macroscopic dynamical laws using non-equilibrium
statistical mechanical tools. In sharp contrast to the more conventional
theories developed for homogeneously  distributed or only weakly biased data, these laws
are found to describe a non-trivial and persistently non-deterministic macroscopic
evolution, and a generalisation error which retains both stochastic and
sample-to-sample fluctuations, even for infinitely large networks.
Furthermore, for the standard error-correcting microscopic algorithms (such as
the perceptron learning rule) one obtains learning curves with
distinct bias-induced phases.
Our theoretical predictions find
excellent confirmation in numerical simulations.
\end{abstract}

\begin{center}
\end{center}

\tableofcontents

\clearpage
\section{Introduction}

Rosenblatt \cite{rosenblatt1962} first introduced
the perceptron and proved the famous perceptron convergence
theorem in 1962. It is an indicator of the richness of the perceptron as a dynamical system that almost 40 years
later it continues to yield fascinating results which have hitherto remained hidden.
Especially during the last decade, considerable progress has been made in understanding the dynamics of
learning in artificial neural networks through the application
of the methods of statistical mechanics.
The dynamics of on-line learning in perceptrons has been analysed intensively,
but for the most part such studies \cite {perceptrononlinelearning} have been carried out in the
idealised scenario of so-called complete training sets (in which
the number of training examples is large compared with $N$, the number of
degrees of freedom), and have also assumed a homogeneous input data distribution.
A recent review of work in this field is contained in
\cite{macecoolen1}. A general theory of learning in the context of restricted training
sets (where the size of the training set is proportional to $N$) is generally much more difficult,
although an exact solution of the dynamical equations
for the more elementary problem of unbiased on-line Hebbian
learning with restricted training sets and noisy teachers
has been found \cite{{raesollichcoolen1},{raesollichcoolen2}}.
Nevertheless, substantial progress has been made towards a general theory
of learning with restricted training sets and the reader may refer, for example,
to \cite{{macecoolen2},{coolensaad},{heimelcoolen},{LiWong}}, for details.

In this paper we consider complete training sets, but we admit
the possibility of a structural bias of the input vectors.
This is a significant issue since in
real-world situations a training sample will generally have a non-zero average;
this is especially important
in the case of on-line learning, where examples are not
available prior to learning, so that one cannot correct
for any bias prior to processing.
This in itself would be sufficient motivation for the present
study. However, it turns out that the introduction of structurally biased input data
leads to qualitative (rather than only quantitative) modifications of
the actual learning curves observed in numerical
simulations and the mathematical theories required for their
description.
Various authors
\cite{{marangisollabiehlriegler}, {opper}}
 have studied so-called clustered examples,
in which examples are drawn from two Gaussian distributions situated close to each other,
with an input bias of order $N^{-\frac{1}{2}}$ (i.e. in magnitude similar
to finite-size effects).   Learning with
input bias has also been considered in  in the context
of linear networks  \cite{sollichbarber3};
the linear theory was then used to construct an approximation for a class of
non-linear models, and it was shown that on-line learning is more robust to input bias and
out-performs batch learning when such bias is present.

\begin{figure}[t]
\vspace*{-3mm}
  \begin{center}
  \begin{tabular}{r@{}c}
  \parbox[b][70mm][c]{3mm}{\large $E_g$}  &
  \hspace*{3mm} \epsfig{file=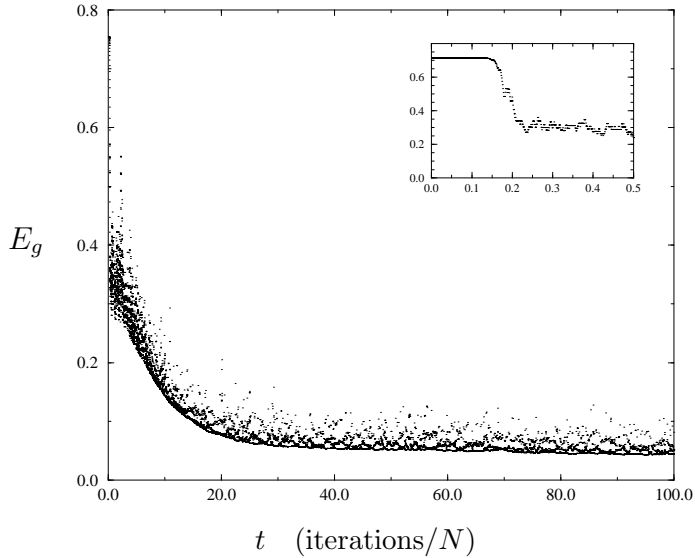, height=70mm}  \\[1mm]
         &   $t~~$  (iterations/$N$)
  \end{tabular}
  \caption{
    Evolution of the generalisation error $E_g$
    as measured in a single simulation experiment of the perceptron rule,
    with $N=1000$, learning rate $\eta=1$ and bias $a=\frac{1}{2}$, following
    initial conditions
    $Q(0)=10$, $R(0)=0$ and $S(0)=\sqrt{N}$ (see the text for details).
    The inset, magnifying the
    early transients, shows the phases I and II. Clearly, no learning
    takes place in phase I.}
    \label{fig:Eg}
\end{center}
\end{figure}

Here we consider a situation which is more natural and less restrictive than the one considered
in \cite{{marangisollabiehlriegler}, {opper}}, and which does not require the
linearity of \cite{sollichbarber3}: we
study the familiar (non-linear) perceptron, with the perceptron learning rule and with a structural,
i.e. $\order(N^0)$,
bias in the input data.  Using $\bB$,
$\bJ$, and $\ba $ to denote the teacher weights, the student
weights and the bias vector (precise definitions follow), we develop our
 theory in terms of three macroscopic  observables:
the standard observables  $Q=\bJ^2,~R=\bJ \cdot \bB,$ and a new observable $S=\bJ \cdot \ba $
(the overlap between student weights and bias vector).
In contrast to the the dynamics of the bias-free case, we find that
in the presence of an $\order(N^0)$ input bias the system passes though three
phases, characterised by different scaling of typical times and of macroscopic observables.
This could already have been anticipated on the basis of numerical
simulations, see e.g. figure \ref{fig:Eg}.
We obtain a closed system of equations in which the evolution of $\{Q,R\}$ is
deterministic in the limit $N\rightarrow \infty $, as in the bias-free case, but
where $S$ is (generally) a stochastic
variable, whose conditional probability distribution $P_t(S|Q,R)$ becomes non-trivial.
Phase I is a short phase, in which the system reduces the alignment of the student
weight vector $\bJ $ relative to the bias vector $\ba$. During phase I, the observable $S$ is
deterministic, is rapidly driven towards zero, and no learning takes place.
Before the state $S=0$ is
reached, however, the system enters phase II, a very short phase
in which $S$ evolves stochastically to a
quasi-stationary probability distribution (which we calculate) and in which both $Q$ and $R$ are frozen.
In phase III, where most of the learning takes place, the $S$ distribution is
modified by a non-negligible random walk element, which generates a diffusion term
in the equation controlling the evolution of
$P_t(S|Q,R)$, whereas $Q$ and $R$ satisfy coupled differential equations which involve
averages over $P_t(S|Q,R)$.
The stochastic nature of $S$ is reflected in the
fact that the generalisation error also exhibits fluctuations (see figure \ref{fig:Eg}).
The (exact) equations describing phase III cannot be further simplified, but we introduce an approximation
 yielding more tractable equations for $\{Q,R\}$, which still have the merit
 of reducing to the more familiar equations when no-bias
is present. Moreover, they are found to be in excellent agreement with the results
obtained from numerical simulations.  Compared to the unbiased case, having a finite bias is found to change
the pre-factor in the asymptotic power law of the asymptotic time-dependence of the generalisation error,
but not the exponent.
A preliminary and more intuitive presentation of some of the present
results can be found in \cite{nips2000}.

\section{Definitions}

We study on-line learning in a student perceptron $\Sigma:\{-1,1\}^N\rightarrow \{-1,1\}$, which learns a task defined by a
teacher perceptron $T:\{-1,1\}^N\rightarrow \{-1,1\}$ whose fixed weight vector is
${\bB } \in {\Re }^N$. Teacher and student output are given by the
familiar recipes
\bd
T(\bxi )=\sgn[{\bB}\cdot\bxi],
~~~~~~~~~~~~
\Sigma(\bxi )=\sgn[{\bJ}\cdot\bxi],
\ed
We assume that $\bB $ is normalised such that $\bB ^2=1$,
the components being drawn randomly with mean zero and standard deviation of order
$\order(N^{-1/2})$, and statistically independent of the input data. In order to model the bias in the
input sample we assume that for $\bxi =(\xi _1,\ldots ,\xi _N)\in \{-1,1\}^N$
all $\xi _i $ are independent,  with $\bra \xi _i\ket =a$, so that
the probability of drawing $\bxi $ is given by
\be p(\bxi )=\prod _i \frac{1}{2}[1+a\xi _i ]
\label{eq:input_distribution}
\ee
We define $\xi _i=a+v_i$, such that the (independent) $v_i$ have mean zero and
variance $\sigma ^2=1\minus a^2$, and the short-hand $\ba =a(1,\ldots ,1)$
(i.e. a vector with all $N$ entries equal to $a$, to be referred to as the `bias vector').
The teacher-bias overlap $\bB \cdot \ba $ is now a random parameter
which is $\order (1)$, since $\bra (\bB \cdot \ba )^2\ket=
\sum _{i,j}a^2\bra B_iB_j\ket =a^2$,
whose distribution will be Gaussian for $N\rightarrow \infty
$, with mean 0 and standard deviation $a$.

The student perceptron $\Sigma$ is being trained according to an
on-line learning rule of the form
$\bJ _{m+1}=\bJ _{m}+\beps_m$, where at each iteration step an
input vector $\bxi_m$ is drawn independently according to
$(\ref{eq:input_distribution})$, and where
\bd
\beps_m=\frac {\eta }{N}\bxi _m \sgn (\bB \cdot \bxi _m){\cal F}[|\bJ _m|, \bJ _m\cdot
\bxi _m, \sgn(\bB \cdot \bxi _m)]
\ed
For Hebbian
learning, for instance, we have
\bd
{\cal F}[J,u,T]=1:~~~~~~~\beps _m =\frac {\eta }{N}\bxi _m \sgn (\bB \cdot \bxi _m)
\ed
whilst the familiar perceptron learning rule is defined by
\be
{\cal F}[J,u,T]=\theta [-uT]:~~~~~~~
\beps _m =\frac {\eta }{2N}\bxi _m [\sgn (\bB \cdot \bxi
_m)-\sgn (\bJ_m \cdot \bxi _m)]
\label{eq:perceptronrule}
\ee
We will derive, from the microscopic stochastic process for the weight vector $\bJ$,
a macroscopic dynamical theory in terms of
the familiar observables $Q=\bJ^2$ and $R=\bJ \cdot \bB$, as well as (in order to obtain closure)
a new observable $S=\bJ \cdot \ba $ measuring the overlap between the vector $\bJ $
and the bias vector. The teacher and student output can then be written
in the form
\bd
\Sigma(\bxi )=\sgn[\lambda _1+x],~~~~~~
T(\bxi )=\sgn [\lambda _2 +y]~~~~~~{\rm with}~~~~~~
\lambda _1=\hat{\bJ}\cdot \ba,
~~~~~~\lambda _2=\bB \cdot \ba,
\ed
with $\hat{\bJ}=\bJ/|\bJ|$, and
where the local fields $\{x,y,z\}$ are defined by $x=\hat {\bJ}\cdot \bv$, $y=\bB \cdot\bv$
and $z=\hat{\ba }\cdot\bv$ (the latter field $z$ will also enter our calculation in due course).
Note that $\lambda_1=S/\sqrt{Q}$. For large $N$, the three fields $\{x,y,z\}$
are zero-average Gaussian random variables, each with variance $\sigma ^2=1\minus a^2$,
and with correlation coefficients given by
\be
\bra xy\ket =\omega \sigma ^2,~~~~~~\bra xz\ket =\sigma ^2 S/|\ba |,~~~~~~\bra yz\ket
=\sigma ^2\lambda _2/|\ba |.
\label{correlationxyz}
\ee
We note that
equation (\ref{correlationxyz}) implies that $z$ will be independent of $(x,y)$ for large $N$
so that
\be
p(x,y,z)=[\sigma \sqrt {2\pi }]^{-1}e^{-z^2/2\sigma ^2}p(x,y),
~~~~~~~~~~~~
p(x,y)=\left[2\pi \sigma ^2 \sqrt {1\minus \omega
^2}\right]^{-1}
e^{-\frac{1}{2}[x^2-2\omega xy+y^2]/\sigma ^2(1-\omega^2)}
\label{eq:joint_dist}
\ee
with $\omega =\hat {\bJ }\cdot \bB=R/\sqrt{Q}$.
It will turn out that most of the averages to appear in this paper, involving
(\ref{eq:joint_dist}) (to be written as $\bra \cdots\ket$), may be expressed
in terms of the function $K(x)={\rm erf}\,(x/\sqrt {2})$.
The generalisation error $E_g=\bra \theta [-(\hat {\bJ }\cdot \bxi )(\bB \cdot \bxi )]\ket$,
for example,
 can be written as
\be
E_g
=\int\!dxdy~p(x,y)\theta[\minus(\lambda_1\plus x)(\lambda_2\plus y)]
=I_1(\lambda_1,\minus\lambda_2,\minus\omega)+I_1(\minus\lambda_1,\lambda_2,\minus\omega)
\label{eq:firstEg}
\ee
where
\bd
I_1(\lambda _1, \lambda _2, \omega )=
\int_{\lambda _1}^\infty\!\! dx\int _{\lambda _2}^\infty\!\! dy~p(x,y)
=\frac {1}{4}\left[1\minus K\left(\!\frac {\lambda
_2}{\sigma}\!\right)\right]
-\frac {1}{2}\int _{\frac {\lambda _2}{\sigma }}^\infty\!\! Dy~
K\left(\!\frac{\lambda_1\minus \omega \sigma y}{\sigma \sqrt {1\minus \omega ^2}}\!\right)
\ed
with the Gaussian measure
$Dy=(2\pi)^{-\frac{1}{2}}e^{-\frac{1}{2}y^2}dy$
(see appendix \ref{app:integrals} for details). This then gives
\be
E_g
=\frac {1}{2}-
\frac {1}{2}\int_{-\frac {\lambda _2}{\sigma }}^\infty\! Dy~
K\biggl (\frac {\lambda _1\plus \omega \sigma y}{\sigma \sqrt{1\minus \omega ^2}}\biggr )
+\frac {1}{2}\int _{\frac {\lambda _2}{\sigma }}^\infty\! Dy~
K\biggl (\frac {\lambda_1 \minus \omega \sigma
y}{\sigma \sqrt{1\minus \omega ^2}}\biggr )
\label{eq:Eg}
\ee
Note that, due to the identity
$\int_0^\infty\! Dy~K(\omega y/\sqrt {1\minus \omega ^2})
=\frac {1}{2}\minus \frac{1}{\pi}\arccos \omega$,
formula (\ref{eq:Eg}) reduces, as it should, to the well known
expression $E_g=\pi^{-1} \arccos \omega$ in the case where
the input bias is zero (i.e. for $a\to 0$).

\section{From Microscopic to Macroscopic Laws}

We now consider the dynamics of the macroscopic observables $\{Q,R,S\}$ in the limit of large $N.$
In the bias-free case, where for large $N$ the fluctuations in the
macroscopic observables are insignificant, this can be done in
a direct and simple way. Here, for $a\neq 0$, the situation is qualitatively
different, since (as will turn out) the fluctuations in $S$ will
no longer vanish, and their distribution will have a strong impact
on the macroscopic laws.
In order to provide a setting for our theory we briefly review a well known procedure \cite{macecoolen1}
which enables us to pass from a discrete to
a continuous time description. We suppose that at time $t$
the probability that the perceptron has undergone precisely $m$ updates is given by the Poisson
distribution $\pi _{m}(t)=\frac{1}{m!}(Nt)^me^{-Nt}$. For large $N$ this will give us
 $t=\frac{m}{N}+\order(N^{-1/2})$,
the usual real-valued time unit, and the uncertainty as to where we are on the
time axis vanishes as $N\rightarrow \infty$.
It is not hard to show that the probability density $p_t(\bJ )$ of finding the vector $\bJ $ at time $t$
satisfies
\bd
\frac {d}{dt}p_t(\bJ)=N\int\! d\bJ^\prime \left\{
\room
\bra \delta [\bJ \minus \bJ^\prime\minus \beps ]\ket_\bxi-\delta [\bJ \minus
\bJ^\prime]\right\}
p_t(\bJ^\prime)
\ed
where, for the perceptron learning rule (\ref{eq:perceptronrule}), the single-step modification
$\beps$ is given by
\bd
\beps =\frac {\eta }{2N}\bxi~[\sgn (\bB \cdot \bxi)-\sgn (\bJ\cdot \bxi)]
\ed
 and where
and $\bra \cdots \ket_\bxi$ denotes the average over all questions $\bxi $ in the
training set $\{-1,1\}^N$. The macroscopic observables $\bOmega =(Q,R,S)$, in turn,  have
the probability density $P_t(\bOmega )=\int\! d\bJ~ p_t(\bJ )\delta [\bOmega \minus \bOmega (\bJ )]$,
which satisfies the macroscopic stochastic equation
\bd
\frac{d}{dt}P_t(\bOmega )=\int\! d\bOmega^\prime ~{\cal W}_t[\bOmega ,\bOmega^\prime]P_t(\bOmega^\prime)
\ed
where
\bd
{\cal W}_t[\bOmega ,\bOmega^\prime]=N\bra ~\bra\delta [\bOmega \minus \bOmega
(\bJ\plus \beps)]\ket_\bxi-\delta [\bOmega \minus \bOmega (\bJ )]~\ket_{\bOmega^\prime,t}
\ed
with
the so-called sub-shell (or conditional) average
$\bra \cdots \ket_{\bOmega^\prime,t}$,  defined as
\bd
\bra f(\bJ )\ket_{\bOmega,t }=\frac {\int\! d\bJ ~p_t(\bJ )\delta [\bOmega
\minus \bOmega (\bJ )]f(\bJ )}{\int\! d\bJ ~p_t(\bJ )\delta [\bOmega \minus \bOmega (\bJ )]}.
\ed
It is possible to make
various assumptions regarding the scaling behaviour of our observables
at time $t=0$, but once this has been specified the scaling
at subsequent times is determined by the dynamics.
We make the natural assumption that $Q(0)=\order (1)$ so that, in accordance with our
assumptions regarding the statistics of $\bB$, we have $R(0)=\order(N^{-1/2})$. We suppose that
$S(0)=\order(N^{1/2})$,
the maximum permitted by the Schwarz inequality.

In this context it is worth remarking that
in the idealised case of zero bias, Hebbian learning
is known to out-perform the perceptron learning rule; but
in the more realistic situation of even moderately biased data the Hebbian rule fails
miserably. For example, if we assume that $S(0)$ is $\order(1)$,and
that $Q,R$ are initially $\order(1),$ it follows from
the learning rule (or from the methods which we apply below to the perceptron learning rule)
that in the initial evolution of the Hebbian system
$dS/d\tau=\eta a^2K(\lambda _2/\sigma),$
where $\tau =Nt,$ so that $S$ rapidly diverges and no learning
takes place; the student vector $\bJ$ cannot break away from its
alignment to the bias vector. We shall show, however, that the perceptron has
no problem coping with extreme initial conditions
such as $S(0)=\order(N^{1/2}),$ and that in due course  effective learning occurs.
The Hebbian example also serves to show that, even if we were to
choose the weaker initial scaling $S(0)=\order(N^{0})$, dependent on the specific choice we make for the
learning rule, the order parameter $S$ might well be driven towards $S=\order(N^{\frac{1}{2}})$
states.

A systematic exploration of the possible scaling
scenarios
reveals the following.\footnote{For brevity we will in this paper only describe
the resulting self-consistent solution, which is indeed perfectly consistent with
the observations
in numerical simulations such as in figure \ref{fig:Eg}.} For the perceptron learning rule and for the initial
scaling conditions  as specified above, the only self-consistent
solution of the macroscopic equations is one describing a
situation where the system passes through three phases
$\{\I,\II,\III\}$
defined by time scales $t=\{\tau N^{-1/2},\tau N^{-1}, \tau \}$, in which
our observables are $\order (1)$ quantities in all three phases, with
the exception of $S$ which is $\order(N^{1/2})$ in phase I. We will write
$S=\widetilde{S}N^{1/2}$
in Phase I, with $\widetilde {S}=\order(N^0)$, and formulate our Phase I equations
in terms of $\widetilde{S}$ rather than $S$. The number of iterations $m$ is related
to the original time $t$ by $m=Nt$ so that the number of iterations up to time $\tau $,
in each of the three phases, is given by $m=\{\tau N^{1/2}, \tau N^0, \tau N\}$.
We incorporate these scaling properties into our equations in each of the three phases, by
working henceforth only with $\order(N^0)$ time units $\tau$ and $\order(N^0)$ observables $\bOmega$,
which satisfy
\be
\frac{d}{d\tau}P_{\tau}(\bOmega)=\int\! d\bOmega^\prime ~{\cal W}_\tau[\bOmega ,\bOmega^\prime]P_\tau(\bOmega^\prime)
\label{eq:newdPOmegadt}
\ee
with
\bd
{\cal W}_\tau[\bOmega ,{\bOmega }']=F_{\I,\II,\III}~
\bra ~\bra
\delta[\bOmega \minus \bOmega (\bJ\plus \beps)]\ket_\bxi-\delta [\bOmega \minus \bOmega (\bJ )]
~\ket_{\bOmega^\prime,t}
~~~~~~~~~~~~~~
\ed
\be
~~~~~~~~~~~~~~
=\frac{F_{\I,\II,\III}}{(2\pi ) ^3}\bigbra \int\! d\hat
{\bOmega }~e^{i\hat {\bOmega }\cdot \bOmega }\biggl \{ \bra e^{-i\hat
{\bOmega } \cdot \bOmega (\bJ +\beps )}\ket_\bxi -e^{-i\hat
{\bOmega }\cdot \bOmega (\bJ )}\biggr \}\bigket_{\bOmega^\prime,t}
\label{eq:newcalW}
\ee
\no
and $F_\I=N^{1/2},~F_{\II}=N^0,~~F_{\III}=N.$
In a subsequent stage it will be convenient to write $\beps =\bk +\bk^\prime$, where
\be
\bk =\frac {\eta }{2N}\ba[\sgn (\bB \cdot \bxi ) \minus \sgn (\bJ \cdot \bxi )],
~~~~~~~~\bk^\prime=\frac {\eta }{2N}\bv[\sgn (\bB \cdot \bxi ) \minus \sgn (\bJ \cdot \bxi
)]
\label{eq:kkprime}
\ee
so that
\be
\beps \cdot \ba =\frac{1}{2}\eta a^2[\sgn (\lambda _2\plus y) \minus
\sgn (\lambda _1 \plus x))]
+\eta a\frac {z}{2\sqrt {N}}[\sgn (\lambda _2\plus y) \minus  \sgn (\lambda _1 \plus x))]
\label{eq:epsdota}
\ee
We are now in a position to discuss the dynamics in each of the three
phases in which different scaling laws apply.

\section{Phase I: Elimination of Bias-Induced Activation}

In Phase I we define the $\order(N^0)$ observables
$\bOmega =(\widetilde{S},Q,R)=(\bJ \cdot \ba /\sqrt {N},Q,R)$ and $F_I=\sqrt {N}.$
Upon expanding the exponential $e^{-i\hat{\bOmega }\cdot \bOmega (\bJ \plus\beps )}$
in powers of $\beps$ we obtain from equation
(\ref{eq:newcalW})
\bd
{\cal W }_\tau[\bOmega ,\bOmega^\prime]
= -\frac {1}{(2\pi )^3}\int\! d\hat{\bOmega }~\bigbra
e^{i\hat {\bOmega }\cdot [\bOmega -\bOmega (\bJ )]}\biggl\{\cdots \biggr \}_{\I}\bigket_{\bOmega^\prime,\tau}
\ed
\no
where
\bd
\biggl\{\cdots \biggr \}_{\I}=i\sum _{i\mu }\bra N^{1/2}\Delta J_i
\frac {\partial\Omega _\mu}{\partial J_i}
\ket_\bxi {\hat {\Omega  }}_\mu +\frac {1}{2} i\sum _{i j \mu }
\bra N^{1/2} \Delta J_i\Delta J_j\frac {\partial^2\Omega_\mu}{\partial J_i\partial J_j}
\ket_\bxi{\hat {\Omega }}_\mu
~~~~~~~~~~~~~~~~~~~
\ed
\bd
~~~~~~~~~~~~~~~~~~~~~~~~~~~~~~~~~
+\frac {1}{2}\sum _{i j \mu \nu }\bra N^{1/2} \Delta J_i\Delta J_j
\frac {\partial \Omega_\mu }{\partial J_i}\frac {\partial \Omega_\nu }{\partial J_j}
\ket_\bxi {\hat {\Omega }}_\mu {\hat {\Omega }}_\nu +\order(N^{-1}).
\ed
A straightforward calculation using equation (\ref{eq:kkprime}) and the
two averages
$\bra \sgn(\lambda _2\plus y)\ket=K(\frac{\lambda _2}{\sigma })$
and $\bra \sgn(\lambda _1\plus x)\ket=K(\frac{\lambda _1}{\sigma })= \sgn (\widetilde{S})$
(which is valid for large $N$ in phase I) now gives
\bd
\biggl\{\cdots \biggr \}_{\I}=\frac {1}{2}i\eta a^2[K(\frac {\lambda _2}{\sigma })-\sgn(\widetilde{S})]{\hat
{\Omega  }}_1+i\eta \widetilde {S}[K(\frac {\lambda _2}{\sigma })\minus \sgn(\widetilde{S})]{\hat
{\Omega }}_2.
\ed
\no
We can now apply equation
(\ref{eq:newdPOmegadt})
to compute the time derivative of the probability density $P_\tau (\bOmega )$.
Note that the sub-shell average $\bra \cdots \ket _{\bOmega^\prime,\tau}$ involves an integration over
all $\bJ $ for which $\bOmega (\bJ )=\bOmega^\prime$ (in a distributional sense) so
 in calculating the relevant integrals
we may effectively replace $\bOmega (\bJ )$ by $\bOmega^\prime$ at appropriate stages.
For example, $\int\! d\hat{\bOmega}~\hat{\Omega}_j e^{i\hat{\bOmega } \cdot [\bOmega -\bOmega (\bJ )]}
=i(2\pi )^3\partial_{j^\prime}\delta[\bOmega \minus \bOmega^\prime]$,
where $\partial_{j^\prime}$ denotes differentiation with respect to $\Omega^\prime_j.$ We now find
for $P_\tau (\widetilde{S},Q,R)$
a Liouville equation
\bd
\frac {d}{d\tau} P_\tau (\widetilde{S},Q,R)=-\frac {\partial
}{\partial\widetilde{S}}\biggl [\frac {\eta
a^2}{2}[K(\frac {\lambda _2}{\sigma })-\sgn(\widetilde{S})]P_\tau (\widetilde {S}, Q,R)\biggr ]
-\frac {\partial }{\partial Q}\biggl [\eta \widetilde{S}[K(\frac {\lambda _2}{\sigma })-\sgn(\widetilde{S})]P_\tau (\widetilde {S}, Q,R)\biggr ]
\ed
with the deterministic solution
$P_\tau (\widetilde {S},Q,R)=\delta[\widetilde {S}\minus \widetilde
{S}(\tau)]~\delta[Q\minus Q(\tau)]~\delta [R\minus R(\tau)]$,
where the actual deterministic trajectory $\{\widetilde {S}(\tau),Q(\tau),R(\tau)\}$ is the solution
of the coupled flow equations
\bd
\frac{d}{d\tau} {\widetilde {S}}=\frac{1}{2}\eta a^2[K(\frac {\lambda _2}{\sigma })\minus \sgn(\widetilde{S})],
~~~~~~~~~
\frac{d}{d\tau} Q=\eta \widetilde{S}[K(\frac {\lambda _2}{\sigma })\minus \sgn(\widetilde{S})],
~~~~~~~~~\frac{d}{d\tau} R=0.
\ed
It follows that
$\widetilde {S}(\tau)={\widetilde {S}}(0)+\frac{1}{2}\eta a^2\tau [K(\lambda _2/\sigma)\minus
\sgn(\widetilde{S})]$. We
see that $\widetilde {S}$ is driven to zero in times $\tau=\tau _{\pm }$ (with $\pm $ referring to the
cases ${\widetilde {S}}_0>0$ and ${\widetilde {S}}_0 <0 $, respectively), which are given by
\bd
\tau_{\pm }=\frac {2|{\widetilde {S}}_0|}{\eta a^2(1\mp K(\frac {\lambda _2}{\sigma
}))}.
\ed
Irrespective of the value of ${\widetilde {S}}_0$, the system seeks to eliminate
any strong alignment of the learning vector $\bJ $ relative to the bias vector $\ba$.
This is clearly confirmed by numerical simulations.
Our equation for $Q$ also readily integrates to give
$Q=Q_0+[\widetilde {S}^2\minus \widetilde {S}_0^2]/a^2$.
We see that the length $J=\sqrt{Q}$ of the student weight vector decreases and
that $J\rightarrow [J_0^2-\widetilde{S}_0^2/a^2]^\frac{1}{2}$ as $\tau \rightarrow \tau _{\pm
}$. Again, this is confirmed by numerical simulations.
The equation $dR/d\tau =0$ implies that $\omega J $ is constant in Phase I.
As can be clearly seen in figure \ref{fig:Eg}, no learning takes place in this phase, since expression (\ref{eq:Eg})
for $E_g$
reduces to $E_g=\frac {1}{2}[1-\sgn(S)K(\lambda _2/\sigma)]$ in
the limit $|\lambda _1|\rightarrow \infty$ (note: $\lambda_1=S/J$).
However, at times $\tau$ approaching $\tau_\pm$ it is no longer valid to
argue that $S$ is $\order (\sqrt {N})$; it is now $\order (N^0)$
and we enter the scaling regime of Phase II.

\section{Phase II: Transition to Error Correction}

As shown in the previous section, $S$ is an $\order(N^0)$ quantity in phase II, and it is also clear
 that $\{Q,R\}$ are $\order(N^0)$ at the start of phase II. In phase II
(and, as we will see, also in phase III) we have to consider the
observables $\bOmega =(S,\bPhi)$, with $\bPhi =(Q,R)$; the
reason for this slight departure from our phase I
terminology will soon become clear. We can now express equation (\ref{eq:newcalW}) as
\bd
{\cal W }_\tau[\bOmega ,\bOmega^\prime]
=F_{\II,\III}\int\! \frac{d\hat{\bOmega }}{(2\pi ) ^3}
~e^{i\hat {\bOmega }\cdot \bOmega }\bigbra
\bra e^{-i\hat {S} S(\bJ\plus\beps)
-i\sum_{\mu=1}^2 \hat{\Phi}_\mu \Phi_\mu(\bJ +\beps )}\ket_\bxi -e^{-i\hat
{\bOmega }\cdot \bOmega (\bJ )}\bigket_{\bOmega^\prime,\tau}
\ed
Here
\bd
\Phi_\mu (\bJ \plus\beps )=\Phi _\mu (\bJ )+\sum _i\Delta J_i\frac {\partial\Phi_\mu }{\partial J_i}
+\frac {1}{2} \sum _{ij}\Delta J_i\Delta J_j\frac {\partial^2 \Phi_\mu }{\partial J_i\partial J_j}
\ed
(this expansion is exact, since $\{Q,R\}$ are quadratic and linear
functions, respectively).
Substituting and expanding the exponential gives
\bd
{\cal W }_\tau[\bOmega ,\bOmega^\prime]
=\bigbra F_{\II,\III}\int\! \frac{d\hat{\bOmega }}{(2\pi ) ^3}~e^{i\hat {\bOmega }\cdot [\bOmega -\bOmega (\bJ )]}
\bra e^{-i\hat {S}\beps \cdot \ba }\minus 1\ket_\bxi -\int\! \frac{d\hat{\bOmega }}{(2\pi ) ^3}
~e^{i\hat {\bOmega }\cdot [\bOmega -\bOmega (\bJ )]}\biggl\{\cdots \biggr \}_{\II,\III}\bigket _{\bOmega^\prime}
\ed
where
\bd
\biggl\{\cdots \biggr \}_{\II,\III}=iF_{\II,\III}\sum _{i\mu }\bra \Delta J_i
\frac {\partial \Phi_\mu}{\partial J_i}e^{-i\hat {S}\beps \cdot \ba }
\ket_\bxi {\hat {\Phi }}_\mu +\frac {1}{2} iF_{\II,\III} \sum _{ij\mu }
\bra \Delta J_i\Delta J_j\frac {\partial^2\Phi_\mu}{\partial J_i\partial J_j}
e^{-i\hat {S}\bk \cdot \ba }\ket_\bxi {\hat {\Phi }}_\mu
\ed
\be
+\frac {1}{2} F_{\II,\III}\sum _{ij\mu\nu }\bra\Delta J_i\Delta J_j
\frac {\partial \Phi_\mu }{\partial J_i}\frac {\partial \Phi_\nu }{\partial J_j}e^{-i\hat {S}\beps \cdot \ba }
\ket_\bxi {\hat {\Phi }}_\mu {\hat {\Phi }}_\nu +\order (N^{-3/2})
\label{eq:WPhaseIIandIII3}
\ee
\no
Note: whereas it is valid to expand $e^{-i\hat {\bPhi }\cdot \bPhi (\bJ \plus\beps )}$ in the
manner just described, we cannot treat $e^{-i\hat {S}S(\bJ \plus\beps )}$ in the same way since
$\sum _i\Delta J_i (\partial S/\partial J_i)=\ba \cdot \beps =\order (N^0)$ in phases II and III.
Equations (\ref{eq:newdPOmegadt},\ref{eq:WPhaseIIandIII3}) form the basis for our
study of Phases II and III.
\vsp

\begin{figure}
\vspace*{-3mm}
  \begin{center} \begin{tabular}{r@{}c}
    \parbox[b][70mm][c]{10mm}{\large\em times} &
    \epsfig{file=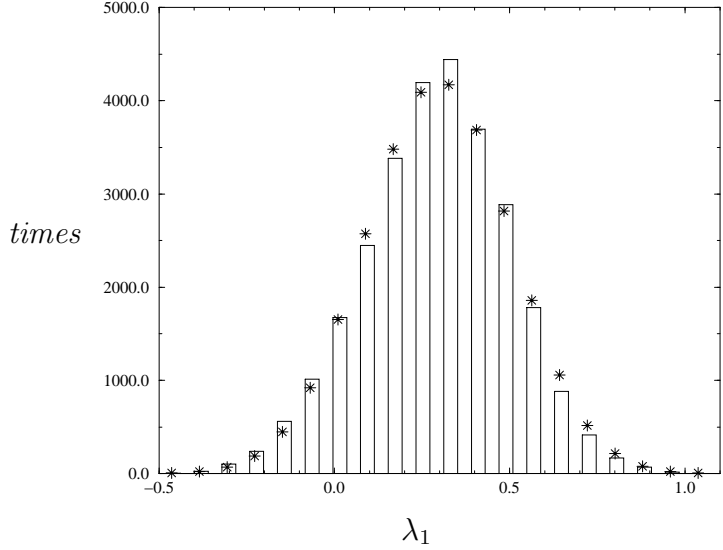, height=70mm} \\[1mm]
    & {\large $\lambda_1$}
  \end{tabular}
  \caption{
  Histogram of values for
  $\lambda_1=S/J$ as measured in  a single simulation experiment with
  $N=400,000$, $\eta=1$,
  and $a=\frac{1}{2}$,
  during the interval $t\in[0, 0.07]$. Here phase I is absent, by virtue of the choice
  $S(0)=0$, and
  $\lambda_2=0.287$. The stars indicate the predicted occurrence probabilities
  as calculated from (\ref{eq:Psolved}). On the short time-scale of observation the
  observed distribution for $\lambda_1$ is truly discrete: no values of $\lambda_1$ were
  found in between the centres of the
  histogram bars.}
   \label{fig:histogram}
  \end{center}
\end{figure}

The time scale $\tau$ in Phase II is related to $t$ via $t=\tau N^{-1}$, so that $F_{\II}=N^0$,
but although this phase is of short duration it has an important role as regards the stochastic
evolution of the  bias overlap parameter $S$.
It is straightforward to show that the third term in equation (\ref{eq:WPhaseIIandIII3})
makes no contribution in the limit of large $N$. Moreover, in the very short phase II we may approximate
$\beps \cdot \ba$ by $\bk \cdot \ba$ (\ref{eq:epsdota}).  Referring to the details and notation in the appendix
we have
\be
\bra e^{-i\hat {S}\bk \cdot \ba }\ket =1-E_g+e^{i\eta a^2\hat {S}}I_1(\minus \lambda_1,\lambda _2,
\minus \omega )+e^{-i\eta a^2\hat {S}}I_1(\lambda_1,\minus \lambda _2, \minus \omega ).
\label{eminusishatkdota}
\ee
and we then find that in phase II
\bd
{\cal W }_\tau[\bOmega ,\bOmega^\prime]=
I_1(\minus \lambda_1, \lambda _2,
\minus \omega )\delta [S\minus S^\prime\plus \eta a^2]\delta [\bPhi \minus
\bPhi^\prime]
+I_1(\lambda_1, \minus \lambda_2, \omega )
\delta[S\minus S^\prime\minus \eta a^2]\delta[\bPhi \minus \bPhi^\prime]
-E_g\delta [\bOmega \minus \bOmega^\prime].
\ed
Substitution into (\ref{eq:newdPOmegadt}) and repetition of the arguments used
for phase I we find that $Q$ and $R$ remain constant in phase II, whilst
the conditional distribution $P_\tau ( {S}|Q,R)$
satisfies
\bd
\frac {d}{d\tau}P_\tau ({S}|Q,R)=
 I_1(\lambda _1({S}^-),\minus \lambda_2 ,\minus\omega)P_\tau ({S}^- |Q,R)
+I_1(\minus \lambda _1( {S}^+),\lambda_2 ,\minus\omega)P_\tau({S}^+|Q,R)
\ed
\bd
~~~~~~~~~~~~~~~~~~~~~~~~~~~~~~~~~~~~~~~~~~~~~~~~~~~~~~~~~~~~~~~~
-E_g({S},Q,R)P_\tau({S}|Q,R)
\ed
where $S^\pm=S\pm \eta a^2$.
The distribution equilibrates, on the relevant time-scale, to a stationary distribution $P(S|Q,R)$ given
as the solution of
\bd
E_g({S},Q,R)P({S}|Q,R)=
 I_1(\lambda _1({S}^-),\minus \lambda_2 ,\minus \omega) P({S}^-|Q,R)
+I_1(\minus\lambda _1({S}^+),\lambda _2 ,\minus\omega ) P({S}^+|Q,R).
\ed
Using relation (\ref{eq:firstEg})
we find that this equilibrium condition can be written as
$A(S)+B(S)=A(S^+)+B(S^-)$,
where $A(S)=I_1(\minus\lambda _1(S),\lambda_2,\minus \omega )P(S|Q,R)$  and
$B(S)=I_1(\lambda _1(S),\minus\lambda _2,\minus\omega )P(S|Q,R)$.
One can easily show by taking Fourier transforms that it  is
satisfied by $B(S)=A(S^+)$, the correctness of which is evident by substitution. In this phase the
permissible values of $S$ are those which differ from some initial value
$S(0)$ by an integral multiple of $\eta a^2$.
Upon writing the allowed values of $S$ as $S_n=S(0)+n\eta a^2$, we immediately obtain
$P(S|Q,R)=\sum_{n=-\infty}^\infty w(S_{n+1}|Q,R)~\delta[S\minus
S_n]$, where
\be
w(S_{n+1}|Q,R)=\frac {I_1(\lambda _1(n),\minus\lambda _2, \minus\omega )}
{I_1(\minus \lambda _1(n\plus 1),\lambda_2, \minus\omega )}~w(S_{n}|Q,R),
\label{eq:Psolved}
\ee
with $I_1$ as given in (\ref{I1}).
Equation (\ref{eq:Psolved}) fully determines the
quasi-stationary distribution $P({S}|Q,R)$. Comparison with numerical simulations shows very
satisfactory agreement, see e.g. figure \ref{fig:histogram}.
The above picture is also in line with our intuition, since in a single step the change in $S$ is given by
\bd
\Delta S=\beps \cdot \ba =\frac{1}{2}\eta a^2[\sgn (\lambda _2\plus y)\minus \sgn (\lambda _1 \plus x)]
+\frac{1}{2} \eta a \frac {z}{\sqrt {N}}[\sgn (\lambda _2\plus y)\minus \sgn (\lambda _1 \plus x)].
\ed
Provided we can neglect the $N^{-\frac{1}{2}}$ term in this expression, which is true
on the time scale of phase II, we see that in a single update $\Delta
S\in\{0,\pm \eta a^2\}$. However, if the $N^{-\frac{1}{2}}$ term could be neglected
indefinitely this would imply that, far into the future, the system would retain a memory of
its initial conditions. In fact
the term $\frac{1}{2}
\eta a z[\sgn (\lambda _2\plus y)\minus \sgn (\lambda _1 \plus x)]/\sqrt{N}$
represents a random walk superposed on the quasi-stationary distribution found for $S$ in
phase II.

\section{Phase III: Error Correction}

As we enter phase III, where $F_{\III}=N$, the above \lq random walk\rq\ term will come to have a significant role after
about $N$ iterations{\footnote{We are grateful to Peter Sollich for pointing this out.},
leading to a modified probability distribution which
contains a diffusion term: $S_n\to S_n+s(t)$. The walk is given by
\bd
s(t)=\frac{\eta a}{2\sqrt{N}}\sum _{\mu =1}^{Nt} z(\mu)[\sgn (\lambda _2\plus y(\mu ))\minus
\sgn (\lambda_1 (\mu )\plus x(\mu ))]
\ed
in an obvious notation,
where the fields $z(\mu)$ are, as we have seen earlier, independent of $(x,y)$. The random walk addition $s(t)$ has mean zero, and variance given by
\be
\bra s^2(t)\ket =\frac {\eta ^2a^2\sigma ^2}{2N}\sum _{\mu =1}^{Nt}
\bra [1\minus \sgn (\lambda _2\plus y(\mu ))\sgn (\lambda _1 (\mu )\plus x(\mu ))]\ket
=t(\eta
a\sigma )^2 \bra E_g\ket
\label{walk}
\ee
where $\bra E_g\ket$ is to be interpreted as a time average of $E_g$
over phase III, up to time $t$.

\begin{figure}[t]
\vspace*{-3mm}
\begin{center}
\begin{tabular}{r@{}c}
\parbox[b][70mm][c]{2mm}{\large $J,~\tilde{S}$}   &
\hspace*{5mm} \epsfig{file=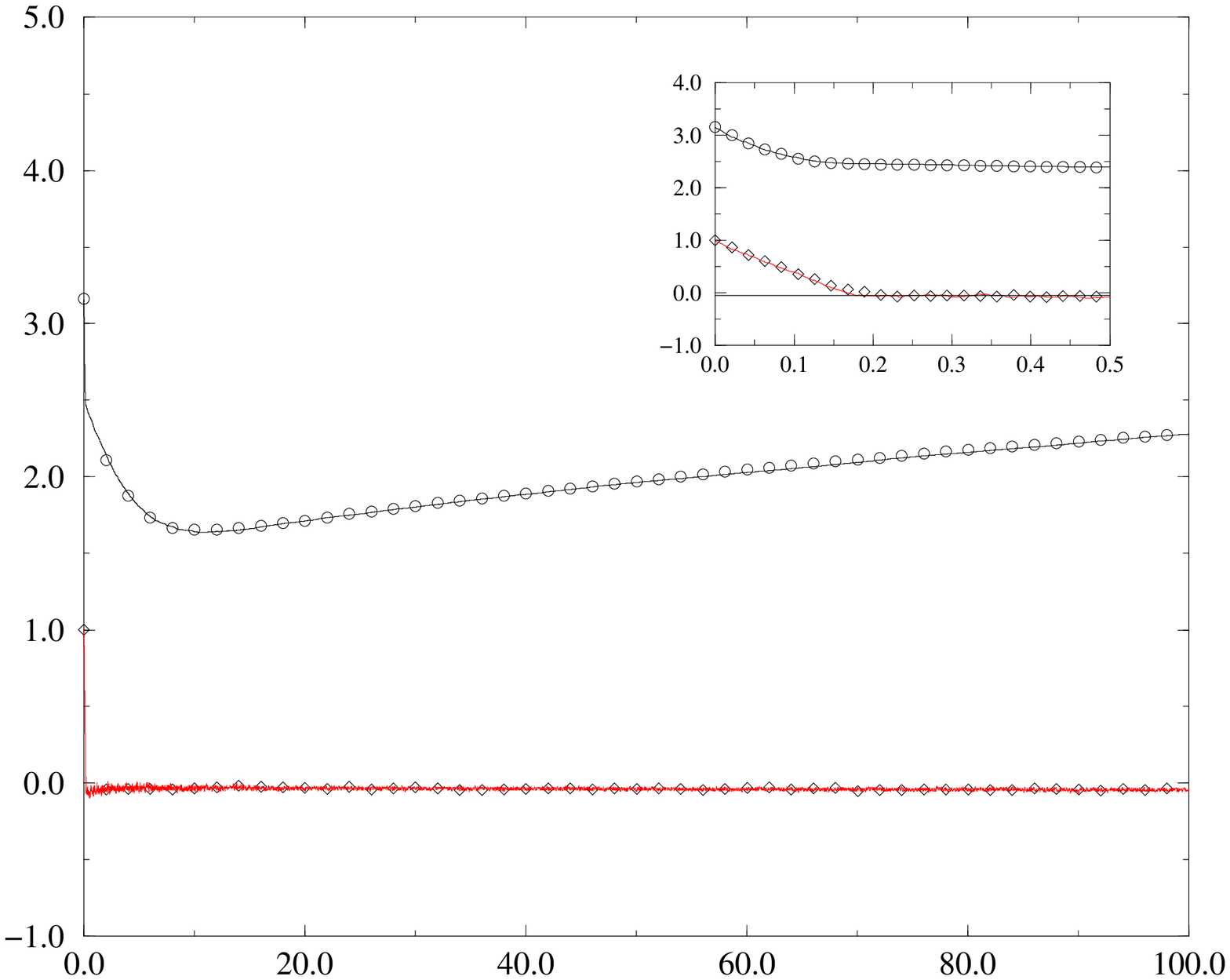, height=70mm} \\
         &   {\large $t$}
\end{tabular}
\caption{
Evolution of the order parameters $J=|\bJ|$ (upper) and
$\tilde{S}=S/\sqrt{N}=\bJ\cdot\ba/\sqrt{N}$ (lower), for $a=\frac{1}{2}$ and $\eta=1$.
Markers indicate simulation results (for $N=1000$), the solid line is the theoretical curve
obtained by numerical solution of equations
(\ref{QdotPhaseIII},\ref{RdotPhaseIII},\ref{PDEforPtauSgivenQR}).
Initial conditions: $Q(0)=10$, $R(0)=0$ and $\tilde{S}(0)=1$. On the time-scale $t=\mu/N$ only phase III is
visible. The inset shows a magnification of the initial stage of the process,
where phase I can be observed.}
\label{fig:JS}
\end{center}
\end{figure}

In order to extract the macroscopic laws in phase III we
will now have to analyse this diffusion effect carefully, starting
from equation (\ref{eq:WPhaseIIandIII3}).
The details of this analysis are given in appendix
\ref{app:phaseIIIdetails}, where we show that for large $N$ the macroscopic
distribution in phase
III will again be of the form  $P_\tau(S,Q,R)=P_\tau(S|Q,R)\delta[Q\minus Q(\tau)]\delta[R\minus
R(\tau)]$,
but now with the deterministic values $\{Q(\tau),R(\tau)\}$ given as the
solution of the coupled equations
\be
\frac{d}{d\tau}Q =\eta \sqrt{Q}\int\! dS~ (K_1\plus L_1\plus M_1)P_\tau (S|Q,R)
+\frac {1}{2}\eta ^2\int\! dS~ (K_3\plus L_3\plus M_3)P_\tau (S|Q,R)
\label{QdotPhaseIII}
\ee
\be
\frac{d}{d\tau}R=\frac {1}{2}\eta \int\! dS~ (K_2\plus L_2\plus M_2)P_\tau (S|Q,R)
\label{RdotPhaseIII}
\ee
The factors $\{K_i,L_i,M_i\}$, defined in appendix
\ref{app:phaseIIIdetails}, are indeed functions of
$S$ (via $\lambda_1$) and of $\{Q,R\}$.
The origin and meaning of these two equations can be appreciated more clearly by writing
them in the following, somewhat more
appealing, form (without as yet specifying the learning rule ${\cal F}[J,u,T])$:
\bd
\frac{d}{d\tau}Q =2\eta J\int\! dS~P_\tau (S|Q,R)~\bra
(\lambda_1\plus x)\sgn(\lambda_2\plus y)
{\cal F}[\sqrt{Q},\lambda_1\plus x,\sgn(\lambda_2\plus y)]\ket
~~~~~~~~~~
\ed
\bd
~~~~~~~~~~~~~~~~~~~~~~~~~~
+\eta ^2\int\! dS~ P_\tau (S|Q,R)~\bra {\cal F}^2[\sqrt{Q},\lambda_1\plus x,\sgn(\lambda_2\plus y)]\ket
\ed
\bd
\frac{d}{d\tau}R=\eta \int\! dS~P_\tau (S|Q,R)~
\bra|y|
{\cal F}[\sqrt{Q},\lambda_1\plus x,\sgn(\lambda_2\plus y)]\ket
\ed
(see \cite{nips2000}  for details).
Although equations (\ref{QdotPhaseIII},\ref{RdotPhaseIII})
are superficially similar to the equations which we derived in
phase I, we now have a situation in which functions of $S$
are weighted with respect to the probability
distribution $P_\tau (S|Q,R)$ which satisfies a partial
differential equation derived from equation (\ref{PDEforPtauSQR})
(in appendix \ref{app:phaseIIIdetails}) by integration over $Q$ and $R$,
namely
\bd
\frac{d}{d\tau}P_\tau (S|Q,R)=\hspace*{130mm}
\ed
\bd
N\left[\room
I_1(\minus \lambda_1(S^+\!,\lambda_2 ,\minus \omega ) P_\tau (S^+|Q,R)
+I_1(\lambda_1(S^-,\minus \lambda_2 ,\minus \omega ) P_\tau(S^-|Q,R)
-E_g(S,Q,R)P_\tau (S|Q,R)\right]
\ed
\be
+\frac {1}{2}\eta ^2a^2\sigma ^2\biggl [
\frac{\partial^2}{\partial S^2}
[I_1(\minus\lambda_1(S^+),\lambda_2,\minus \omega )P_\tau(S^+|Q,R)]
+\frac{\partial^2}{\partial S^2}[I_1(\lambda_1(S^-), \minus\lambda_2, \minus\omega )
P_\tau (S^-|Q,R)]
\biggr ]
\label{PDEforPtauSgivenQR}
\ee
Equations (\ref{QdotPhaseIII},\ref{RdotPhaseIII},\ref{PDEforPtauSgivenQR}), together with the
definitions of the short-hands $\{K_i,L_i,M_i\}$ as given in appendix
\ref{app:phaseIIIdetails},
 provide an exact and
closed set of equations for the macroscopic dynamics in phase III, in terms of the  observables $\{S,Q,R\}$.
In the large $N$ limit,
$Q$ and $R$ satisfy deterministic equations, as in conventional no-bias theories,
but $S$ remains stochastic throughout phase III. Furthermore, the persistent appearance of the factor $\lambda_2$
(which depends on the actual realisation of the teacher weights) induces sample-to-sample fluctuations.
An example of the result of solving the coupled equations (\ref{QdotPhaseIII},\ref{RdotPhaseIII},\ref{PDEforPtauSgivenQR})
numerically (via a numerical realisation, i.e. Monte Carlo, of the conditional stochastic process (\ref{PDEforPtauSgivenQR}) for $S$)
is shown in figure \ref{fig:JS}, and compared with
numerical simulations of the underlying microscopic perceptron learning
process. The agreement between theory and experiment is quite
satisfactory.

\section{Asymptotics of the Generalisation Error}

 A full numerical study of our equations (\ref{QdotPhaseIII},\ref{RdotPhaseIII},\ref{PDEforPtauSgivenQR}) would
be difficult, but these equations undergo a great simplification,
permitting further analysis, if we make the approximation  $P_\tau (S|Q,R) =\delta [S\minus \bra S\ket
]$, and assume that $\lambda_1 (\bra S \ket)=\lambda_2$;
numerical simulations confirm the validity of the replacement of $\lambda _1$
by $\lambda _2$ {\em {on average}} in phase III. In this approximation equations
(\ref{QdotPhaseIII},\ref{RdotPhaseIII}) become
\bd
\frac{d}{d\tau} Q=\eta \sqrt{Q}[K_1\plus L_1\plus M_1]+\frac{1}{2}\eta ^2[K_3\plus L_3\plus M_3]
~~~~~~~~~~~~~~~
\frac{d}{d\tau} R=\frac {1}{2}[K_2\plus L_2\plus M_2]
\ed
Note that
$K_1\plus L_1\plus M_1=\lambda _1[(A_1\plus B_1\plus C_1)\minus (A_2\plus B_2\plus C_2)]\plus
(A_3\plus B_3\plus C_3)\minus (A_4\plus B_4\plus C_4)$.
Referring to appendix \ref{app:integrals} for the relevant expressions for $\{A_i,B_i,C_i\}$
in terms of the integrals
$I_1(\lambda _1,\lambda _2,\omega )$ and $I_2(\lambda _1,\lambda _2,\omega )$, and using the identity
$K(\alpha )=\int _{-\infty }^\infty~ Dy~K((\alpha \minus \omega y)/\sqrt{1\minus
\omega^2})$,
we find that in the approximation $\lambda _1=\lambda _2$ the
following identities hold:
\bd
A_1\plus B_1\plus C_1=K(\lambda _2/\sigma ),~~~~~~A_2\plus B_2\plus C_2=K(\lambda _2/\sigma )
\ed
\bd
A_3\plus B_3\plus C_3=\sqrt {\frac {2}{\pi }}\omega \sigma e^{-\frac {\lambda _2^2}{2\sigma ^2}},
~~~~~~A_4\plus B_4\plus C_4=\sqrt {\frac {2}{\pi }}\sigma e^{-\frac {\lambda _2 ^2}{2\sigma ^2}},
~~~~~~K_1\plus L_1\plus M_1=\minus \sqrt {\frac {2}{\pi }}\sigma (1\minus \omega )
e^{-\frac {\lambda _2 ^2}{2\sigma ^2}}.
\ed
\bd
K_2\plus L_2\plus M_2=(A_5\plus B_5\plus C_5)-(A_6\plus B_6\plus C_6)= \sqrt {\frac {2}{\pi }}\sigma (1\minus \omega )
e^{-\frac {\lambda _2^2}{2\sigma ^2}}
\ed
\bd
K_3\plus L_3\plus M_3=1-\int _{-\frac {\lambda _2}{\sigma }}^\infty Dy\,K\biggl
(\frac {\lambda _2\plus \omega \sigma y}{\sigma \sqrt {1\minus \omega ^2}}\biggr )
+\int_{\frac {\lambda _2}{\sigma }}^\infty Dy\,K\biggl
(\frac {\lambda _2\minus \omega \sigma y}{\sigma \sqrt {1\minus \omega ^2}}\biggr
)=2E_g
\ed
and equations (\ref{QdotPhaseIII},\ref{RdotPhaseIII})  therefore
become (upon rewriting the equation for $Q$ in terms of $J=\sqrt{Q}$):
\be
\frac{d}{d\tau}J=- \frac {\eta}{\sqrt{2\pi }}\sigma
(1\minus \omega )
e^{-\frac {\lambda _2 ^2}{2\sigma ^2}}
+\frac {\eta ^2}{2J}E_g
~~~~~~~~~~~~~~~
\frac{d}{d\tau} R= \frac{\eta}{\sqrt{2\pi }}\sigma (1\minus \omega )e^{-\frac {\lambda _2^2}{2\sigma ^2}}
\label{dJdRdtfinal}
\ee
The corresponding equation for $\omega=R/J$ is
\be
\frac {d}{d\tau}\omega =\frac {\eta}{J\sqrt{2\pi
}}\sigma (1\minus \omega
^2)e^{-\frac {\lambda _2^2}{2\sigma ^2}}
-\frac {\omega \eta ^2}{2J}E_g
\label{domegadtfinal}
\ee
which is to be solved in combination with (\ref{eq:Eg}).
Numerical solution of these equations is found to be in very good agreement
with the results of numerical simulations, even for finite times; however,
it is relevant to consider what basis exists for making the
approximation $\lambda_1=\lambda_2$, other than the fact that it works. We have already
observed that the probability distribution for $S$ in phase III is a
random walk superposed on the underlying discrete distribution which emerged in phase
II. Equation (\ref{walk}) indicates
that the random walk, reflected in the diffusion
terms in equation (\ref{PDEforPtauSgivenQR}), could in principle
lead to a large variance for $S$, were this random walk not coupled to the
underlying discrete distribution via equation (\ref{PDEforPtauSgivenQR}).
The discrete distribution and the random walk, however, are found to interact in such a
way that the fluctuations actually tend to zero in the limit
$\tau\to
\infty$; this is confirmed by the results of numerical simulations which show that the fluctuations in
$\lambda _1=S/J$ decrease with time and that on average $\lambda _1$ tends to $\lambda
_2$, see figure \ref{fig:longhistogram}.
In a single step the {\it average change} in $S$ is equal to
\bd
\frac{1}{2}\eta a^2\bra [\sgn(\lambda _2\plus y)\minus \sgn
(\lambda _1\plus x)]\ket +\frac{1}{2}\eta a \frac{\bra z\ket}{\sqrt
{N}}\bra [\sgn(\lambda _2\plus y)\minus \sgn
(\lambda _1\plus x)]\ket
=\frac{1}{2}\eta a^2[K (\frac{\lambda
_2}{\sigma })-K(\frac {\lambda _1}{\sigma })]
\ed
so, as the fluctuations in $S$ diminish, we do indeed expect that $\lambda_1$
will tend to $\lambda_2.$

\begin{figure}[t]
\vspace*{-3mm}
  \begin{center} \begin{tabular}{r@{}c}
    \parbox[b][70mm][c]{3mm}{\large $p(\lambda_1)$}&
  \hspace*{3mm}  \epsfig{file=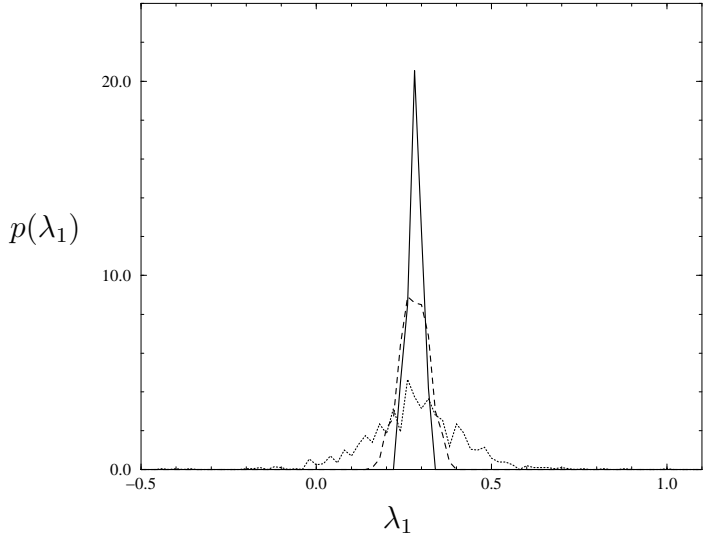, height=70mm} \hspace*{3mm}\\
    & {\large $\lambda_1$}
  \end{tabular}
  \caption{
Distribution $p(\lambda_1)$ of $\lambda_1=S/J$, as measured during a single simulation
run, with $N=1000$, $a=\frac{1}{2}$ and $\eta=1$,
over the time intervals
$[0,100]$ (dotted curve), $[950,1050]$ (dashed curve) and $[9900,10000]$
(full curve). One observes that the fluctuations in $\lambda_1$
are reduced to zero, as time progresses.
  }
\label{fig:longhistogram}
\end{center}
\end{figure}

We will now use the coupled equations (\ref{eq:Eg},\ref{domegadtfinal})
to derive an asymptotic expression for the
generalisation error $E_g$. Differentiation of (\ref{eq:Eg}) with respect to $\omega$
gives
\bd
\frac {\partial E_g}{\partial \omega }=-\frac {e^{-\frac {1}{2}\beta^2}}{\pi \sqrt {1\minus \omega
^2}}
e^{-\frac {1}{2}\beta ^2(1\minus \omega)/(1\plus \omega)}
\ed
with the constant $\beta =\lambda _2/\sigma$.
Changing the variable to $\omega =\cos \theta $, and expanding for $\theta \rightarrow 0$ gives
\be
E_g=\pi^{-1} e^{-\frac {1}{2}\beta ^2}\int _0^\theta\! du~ e^{-\frac {1}{2}\beta ^2
\tan ^2(u/2)}
= \pi^{-1} e^{-\frac {1}{2}\beta ^2} [\theta -\frac {\beta ^2\theta
^3}{24}+\order(\theta ^5)]
\label{asymptoticEg}
\ee
Equation (\ref{dJdRdtfinal}) for $J$ and equation (\ref{domegadtfinal}) for $\omega $ can now be written
\bd
\frac {dJ}{d\tau}=-\frac {\eta \sigma}{\sqrt {2\pi }}(1\minus \cos \theta )e^{-\frac{1}{2}\beta
^2}\!
+\frac{\eta^2 E_g}{2J}
~~~~~~~~~~~
-J\sin \theta \frac {d\theta }{d\tau }=\frac {\eta \sigma }{\sqrt
{2\pi }}\sin ^2\theta ~e^{-\frac{1}{2}\beta ^2}\!-\frac {\eta ^2 E_g \cos \theta
}{2J}
\ed
Using the expansion
$\tan \theta = \theta +\frac{1}{3}\theta^3+\order(\theta ^5)$
we then expand our previous equations for the evolution of $J$ and
$\theta$, giving
\bd
\frac {d}{d\tau}\theta=e^{-\frac{1}{2}\beta
^2}\left\{
-\frac{\eta \sigma \theta }{J\sqrt {2\pi }}+\frac {\eta ^2}{2\pi J^2}-\frac{\eta
^2\theta ^2\rho}{2\pi J^2}\right\} +\order (\theta ^4),~~~~~~~~{\rm with}~~~~~~~~\rho
=\frac{1}{24}\beta ^2+\frac{1}{3}
\ed
\bd
\frac{d}{d\tau}J=e^{-\frac{1}{2}\beta
^2}\left\{
-\frac{\eta \sigma \theta ^2}{2\sqrt {2\pi }}
+\frac {\eta ^2}{2\pi J}[\theta -\frac{1}{24}\beta ^2\theta^3]
\right\}+\order(\theta ^5).
\ed
Upon making the asymptotic ansatz $J=A/\theta$, the equation for $J$ can
now be expressed so as to give a second equation for $\theta$.
The two resulting equations for $d\theta/d\tau$ are
\bd
\frac{d}{d\tau}\theta=\frac{\eta \sigma \theta ^4}{2A\sqrt {2\pi
}}e^{-\frac{1}{2}\beta ^2}-\frac {\eta ^2\theta ^4}{2A^2\pi
}\biggl [1-\frac {\beta ^2\theta ^2}{24}+\order(\theta ^4)\biggr ]e^{-\frac{1}{2}\beta ^2}
\ed
\no
and
\bd
\frac{d}{d\tau}\theta =-\frac{\eta \sigma \theta ^2}{A\sqrt{2\pi
}}e^{-\frac{1}{2}\beta ^2}+\frac{\eta ^2\theta ^2}{2\pi A^2}e^{-\frac{1}{2}\beta
^2}+\order(\theta ^4)
\ed
Consistency requires that $A$ be given by $A=\eta/\sigma \sqrt{2\pi
}$. The asymptotic equation for $\theta $ subsequently becomes
$d\theta /d\tau =-\frac{1}{2}\sigma ^2\theta ^4e^{-\frac{1}{2}\beta
^2}$,
from which we obtain the asymptotic power law $\theta =k\tau ^\alpha ,$
where $\alpha =-\frac{1}{3}$ and $k^3=2e^{\frac{1}{2}\beta^2}/3\sigma^2$.
Combining this, finally, with (\ref{asymptoticEg}) we then obtain,
recalling that in phase III one simply has $\tau=m/N=t$:
\be
E_g(t)=\rho(a) e^{- \lambda _2^2/3\sigma ^2} t^{-\frac{1}{3}}
~~~~(t\to \infty),~~~~~~~~~~~~~
  \rho(a)=\left[\frac{2}{3\pi^3} \right]^{\frac{1}{3}}
  (1\minus a^2)^{-\frac{1}{3}}
\label{eq:egasymp}
\ee
Note that the power of $\tau $ occurring in this
expression is the same as the power which appears in the
asymptotic form of the generalisation error in the conventional
no-bias theory; the  coefficient is however different, but reduces to the familiar form in the
case of zero bias, where $a=\lambda _2 =0$ and $\sigma =1$. Moreover, our prediction of the asymptotic
form of $E_g$ is in excellent agreement with the results of
numerical simulations. This is evident from figure
(\ref{fig:asympcoeff}), where we show the observed function
$\rho(a)$, defined as $\rho(a)=\lim_{t\to\infty} E_g(t)t^{\frac{1}{3}}e^{\lambda_2^2/3\sigma
^2}$,
versus the theoretical prediction as given in (\ref{eq:egasymp}).
Note that
the dependence of (\ref{eq:egasymp}) on the teacher-bias
overlap $\lambda_2=\bB\cdot\ba$ implies sample-to-sample
fluctuations.

\begin{figure}
  \begin{center} \begin{tabular}{r@{}c}
    \parbox[b][70mm][c]{3mm}{\large $\rho(a)$}&
    \epsfig{file=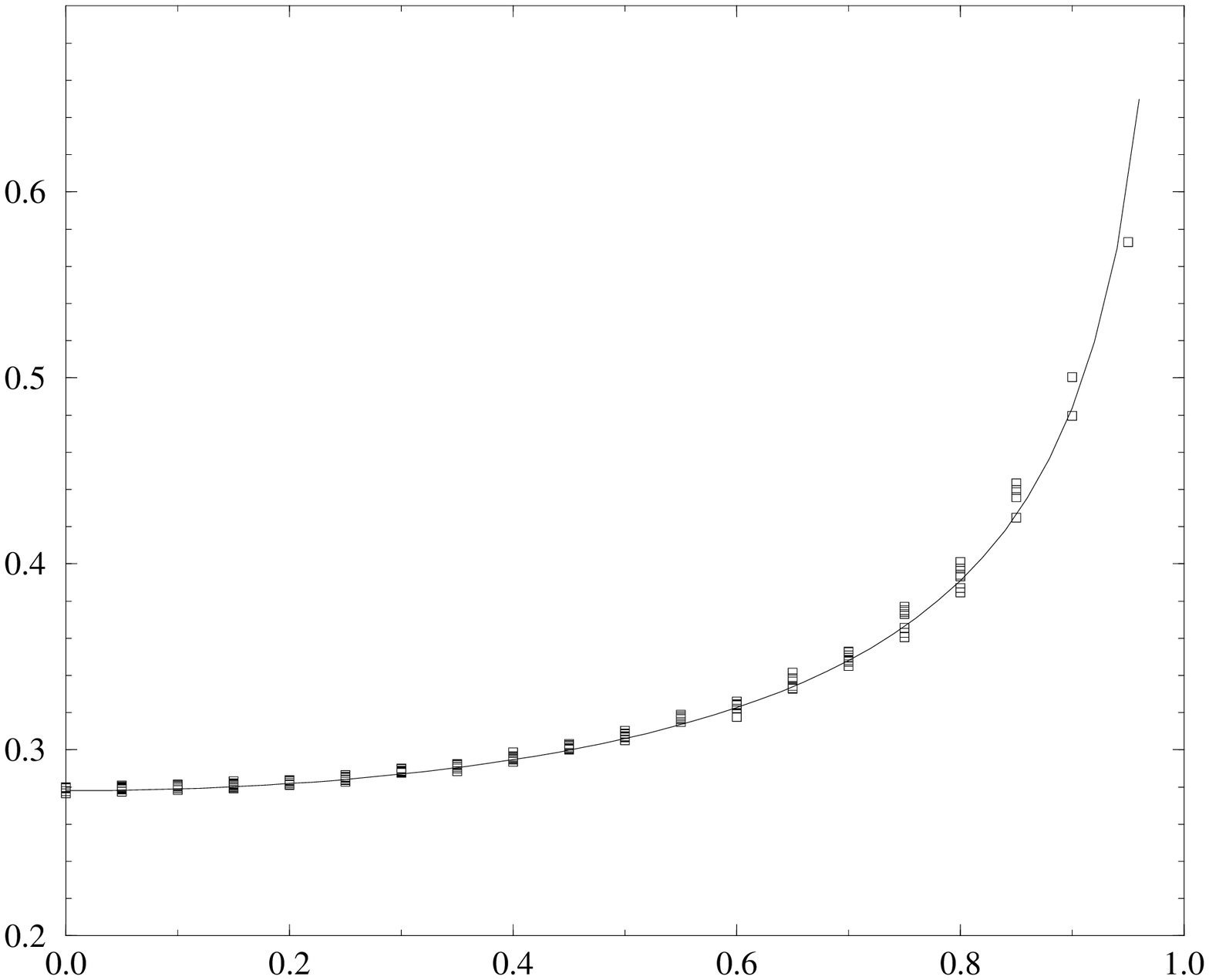, height=70mm} \\
    & {\large $a$}
  \end{tabular}
  \caption{
  Comparison of $\rho(a)$ as found in
simulations ($N=1000$ and $\eta=1$, see the main text for details of its definition),
for various values of the teacher-bias overlap $\lambda_2=\bB\cdot\ba$ (squares), with the
theoretical prediction  (\ref{eq:egasymp}) (solid curve). }
\label{fig:asympcoeff}
\end{center}
\end{figure}

\section{Discussion }

We have studied analytically the dynamics of on-line learning in
non-linear perceptrons, trained according to the perceptron rule, for the scenario of having
structurally biased, i.e. $\order(N^0)$, input data. The bias changes qualitatively the
learning process, inducing three distinct phases (with different scaling properties)
and persistent stochastic as well as sample-to-sample fluctuations in the generalisation error,
even for $N\rightarrow \infty$.
At a theoretical level, the need to introduce an extra order
parameter $S$ (the projection of the student weight vector in the direction of the bias)
which is neither deterministic nor self-averaging makes
the analysis considerably more involved than that of the idealised
bias free case. In the third and final phase, in which most of the
learning takes place, we have obtained a set of exact closed equations
which involve the conditional probability density of $S$. However, because
of their complicated nature, an exact analytic solution of these equations appears to
be out of the question, as is also generally the case in the more familiar no-bias scenarios.
Nevertheless we have found that an approximate (and much simpler)
version of our equations yields results which are in excellent
agreement with numerical simulations.  We show that the asymptotic
power law for the generalisation error is largely preserved, with
the bias showing up only in the pre-factor.
At various stages throughout out calculations we have compared
the predictions of our macroscopic dynamic equations with the
results of numerical simulations of the underlying (microscopic) learning
process, which consistently showed excellent agreement.

Although in this paper we have confined ourselves to the
perceptron learning rule, it is clear that our analysis is in no
way restricted to this particular  rule, and can be applied to other rules such
as the AdaTron learning rule, where $\beps =\frac {\eta }{2N}\bxi
[\sgn (\bB \cdot \bxi ) \minus \sgn (\bJ \cdot \bxi )]|\bJ \cdot \bxi |$;
one could even study optimal
learning rates and optimal learning rules, generalising \cite{KinouchiCaticha}
to the case of having $a\neq 0$.
Preliminary studies of the AdaTron learning rule with
structurally biased data show, for instance, that the simple result (\ref{eq:Psolved}),
describing the phase II distribution in the case of the
perceptron,
is replaced by the integral equation
\bd
E_gP_\tau (\widetilde{S}|Q,R)
=\eta a^2J\int_{-\lambda _1}^\infty\! d\rho~G(\rho, \lambda_2)
~P_\tau (\widetilde{S}\plus(\lambda _1\plus\rho )\eta a^2J)|Q,R)
\ed
\bd
+~\eta a^2J\int _{\lambda _1}^\infty\! d\rho~G(\rho,\minus \lambda_2)
P_\tau (\widetilde{S}\plus(\lambda _1\minus \rho )\eta a^2J)|Q,R).
\ed
where $G$ is defined by
\bd
G(x,\lambda _2)=\frac{e^{-\frac{x^2}{2\sigma ^2}}}{2\sigma\sqrt {2\pi }}
\biggl [1-K\biggl (\frac{\lambda _2\plus \omega x}{\sigma \sqrt {1\minus \omega
^2}}\biggr )\biggr],
\ed
The discrete distribution which in the present paper we found for the perceptron in
phase II no longer applies in the Adatron case, and is replaced by a continuous
distribution which satisfies the above integral equation. The analysis
of the AdaTron in the case of biased data is more complicated than for
the perceptron, as might have been
expected from the nature of the AdaTron learning rule, but much of the work
which we have presented for the perceptron can be carried through and the results
will be published in \cite{raeheimelcoolen02}.
There is also scope for a more detailed mathematical investigation of the partial
differential equation which we derived to describe the conditional
probability distribution $P_\tau (S|Q,R)$ for the perceptron, but this is likely to
be difficult, and beyond the scope of the present paper.

\subsection*{Acknowledgements}

It is a pleasure to thank Peter Sollich and Michael Biehl for
helpful comments and discussions. JAFH wishes to thank King's
College London Association for financial support.

\appendix
\section{Integrals and Averages}
\label{app:integrals}

We recall that the function $K$ is defined by
$K(x)={\rm erf}(x/\sqrt {2})$.
In terms of this definition note that
$\int _{\tau }^\infty\! d\zeta ~e^{-\frac {1}{2}\zeta ^2}=\frac
{1}{2}\sqrt {2\pi }[1\minus K(\tau )]$.
We now proceed to list various integrals which occur in our
calculations, or are referred to in the text, and where
appropriate outline a brief derivation.
Recall that the joint distribution of $(x,y)=(\hat {\bJ } \cdot \bv ,\bB \cdot \bv)$ is given by
\bd
p(x,y)=[2\pi \sigma ^2 \sqrt {1\minus \omega
^2}]^{-1}e^{-\frac{1}{2}\frac {[x^2-2\omega xy+y^2]}{\sigma ^2(1-\omega ^2)}}
\ed
where $\sigma^2=1\minus a^2$ and $\omega =\bB \cdot \hat {\bJ }$.
We then find that
\bd
I_1(\lambda _1, \lambda _2, \omega )=
\int_{\lambda _1}^\infty\!\! dx\int _{\lambda _2}^\infty\!\! dy~p(x,y)
=\int _{\lambda _2}^\infty\!\!\frac{ dy}{2\pi \sigma }~e^{-\frac {y^2}{2\sigma
^2}}\int _{\frac {\lambda _1-\omega y}{\sigma \sqrt {1-\omega
^2}}}^\infty\!\! d\zeta~e^{-\frac {1}{2}\zeta ^2}
\ed
\be
=\frac {1}{4}\left[1\minus K\left(\!\frac {\lambda
_2}{\sigma}\!\right)\right]
-\frac {1}{2}\int _{\frac {\lambda _2}{\sigma }}^\infty\!\! Dy~
K\left(\!\frac{\lambda_1\minus \omega \sigma y}{\sigma \sqrt {1\minus \omega ^2}}\!\right)
\label{I1}
\ee
and similarly
\bd
I_2(\lambda _1 ,\lambda _2, \omega )=\int _{\lambda _1}^\infty\!\!
dx~x\int _{\lambda _2}^\infty\!\!dy~p(x,y)
=\int _{\lambda _1}^\infty\!\! \frac{dx~x}{2\sigma\sqrt {2\pi }}
e^{-\frac {x^2}{2\sigma ^2}}
-\int _{\lambda _1}^\infty\!\! \frac{dx~x}{2\sqrt {2\pi }\sigma }
e^{-\frac {x^2}{2\sigma
^2}}K\left(\!\frac {\lambda _2\minus \omega x}{\sigma \sqrt {1\minus \omega
^2}}\!\right)
\ed
\bd
=\frac{\sigma }{2\sqrt {2\pi }}e^{-\frac {\lambda _1^2}{2\sigma
^2}}\biggl [1\minus K\biggl (\frac {\lambda _2\minus \omega \lambda _1}{\sigma \sqrt {1\minus \omega
^2}}\biggr )\biggr ]
+\frac{\omega \sigma }{2\sqrt {2\pi }}e^{-\frac {\lambda _2^2}{2\sigma
^2}}\biggl [1\minus K\biggl (\frac {\lambda _1\minus \omega \lambda _2}{\sigma \sqrt {1\minus \omega
^2}}\biggr )\biggr ]
\ed
The following averages with respect to the distribution $p(x,y)$ are easily
calculated:
\bd
\bra \sgn (\lambda
_1\plus x)\ket=K\biggl (\frac {\lambda _1}{\sigma }\biggr ),~~~~\bra x~\sgn (\lambda
_2\plus y)\ket =\sqrt {\frac {2}{\pi }}\omega \sigma e^{-\frac {\lambda
_2^2}{2\sigma ^2}},~~~~\bra x~\sgn (\lambda
_1\plus x)\ket =\sqrt {\frac {2}{\pi }}\sigma e^{-\frac {\lambda _1
^2}{2\sigma ^2}}
\ed
\bd
\bra \sgn (\lambda _1\plus x)\sgn
(\lambda _2\plus y)\ket
=I_1(\lambda _1, \lambda _2, \omega )-I_1(\minus \lambda _1, \lambda _2, \minus \omega)
-I_1(\lambda _1, \minus\lambda _2, \minus\omega )+I_1(\minus\lambda _1, \minus\lambda _2, \omega )
\ed
\bd
=\int_{-\frac {\lambda _2}{\sigma }}^\infty\! Dy~K\biggl (\frac
{\lambda _1\plus \omega \sigma y}{\sigma \sqrt {1\minus\omega ^2}}\biggr
)-\int_{\frac {\lambda _2 }{\sigma }}^\infty\! Dy~K\biggl (\frac
{\lambda _1\minus \omega \sigma y}{\sigma \sqrt {1\minus\omega ^2}}\biggr
)
\ed
Finally, in studying phases II and III
we require the following averages:
\bd
\bra \sgn (\lambda _2 \plus y)e^{-i\hat {S}\bk \cdot \ba }\ket
=A_1\plus B_1e^{i\hat {S}\eta a^2}\plus C_1e^{-i\hat {S}\eta a^2}
\ed
\bd
\bra \sgn (\lambda _1 \plus x)e^{-i\hat {S}\bk \cdot \ba }\ket=A_2\plus B_2
e^{i\hat {S}\eta a^2}\plus C_2e^{-i\hat {S}\eta a^2}
\ed
\bd
\bra x~\sgn (\lambda _2 \plus y)e^{-i\hat {S}\bk \cdot \ba }
\ket=A_3\plus B_3e^{i\hat {S}\eta a^2}\plus C_3e^{-i\hat {S}\eta a^2}
\ed
\bd
\bra x~\sgn (\lambda _1 \plus x)e^{-i\hat {S}\bk
\cdot \ba }\ket=A_4\plus B_4e^{i\hat {S}\eta a^2}\plus C_4e^{-i\hat {S}\eta a^2}
\ed
\bd
\bra y~\sgn (\lambda _2 \plus y)e^{-i\hat {S}\bk \cdot \ba }
\ket=A_5\plus B_5e^{i\hat{S}\eta a^2}\plus C_5e^{-i\hat {S}\eta a^2}
\ed
\bd
\bra y~\sgn (\lambda _1 \plus x)e^{-i\hat {S}\bk \cdot \ba }\ket=
A_6\plus B_6e^{i\hat {S}\eta a^2}\plus C_6e^{-i\hat {S}\eta a^2}
\ed
\bd
\bra e^{-i\hat {S}\bk \cdot \ba }\ket =A_7\plus B_7e^{i\hat {S}\eta a^2}\plus C_7e^{-i\hat {S}\eta a^2}
\ed
\bd
\bra \sgn (\lambda _1 \plus x)\sgn (\lambda _2 \plus y)e^{-i\hat {S}\bk \cdot \ba }\ket
=A_8\plus B_8e^{i\hat {S}\eta a^2}\plus C_8e^{-i\hat {S}\eta a^2}
\ed
\clearpage\noindent
where
\bd
\begin{array}{lll}
A_1=-[I_1(\lambda_1,\lambda _2, \omega )-I_1(\minus\lambda _1,\minus\lambda _2, \omega)],
&
B_1=-I_1(\minus\lambda _1,\lambda _2 \minus\omega ),
&
C_1=I_1(\lambda _1,\minus\lambda _2, \minus\omega ),
\\[1mm]
A_2=-[I_1(\lambda _1,\lambda _2, \omega )-I_1(\minus\lambda _1,\minus\lambda _2, \omega )],
&
B_2=I_1(\minus\lambda _1,\lambda _2, \minus\omega ),
&
C_2=-I_1(\lambda _1,\minus\lambda _2, \minus\omega ),
\\[1mm]
A_3=[I_2(\lambda _1,\lambda _2, \omega )+I_2(\minus\lambda _1,\minus\lambda _2, \omega )],
&
B_3=-I_2(\minus\lambda _1,\lambda _2, \minus\omega ),
&
C_3=-I_2(\lambda _1,\minus\lambda _2, \minus\omega ),
\\[1mm]
A_4=[I_2(\lambda _1,\lambda _2, \omega )+I_2(\minus\lambda _1,\minus\lambda _2, \omega )],
&
B_4=-I_2(\minus\lambda _1,\lambda _2, \minus\omega ),
&
C_4=I_2(\lambda _1,\minus\lambda _2, \minus\omega ),
\\[1mm]
A_5=[I_2(\lambda _2,\lambda _1, \omega )+I_2(\minus\lambda _2,\minus\lambda _1, \omega )],
&
B_5=I_2(\lambda _2,\minus\lambda _1, \minus\omega ),
&
C_5=I_2(\minus\lambda _2,\lambda_1, \minus\omega ),
\\[1mm]
A_6=[I_2(\lambda _2,\lambda _1, \omega )+I_2(\minus\lambda _2,\minus\lambda _1, \omega )],
&
B_6=-I_2(\lambda _2,\minus\lambda _1, \minus\omega ),
&
C_6=-I_2(\minus\lambda _2,\lambda _1, \minus\omega ),
\\[1mm]
A_7=1-E_g,
&
B_7=I_1(\minus\lambda _1,\lambda _2,\minus\omega ),
&
C_7=I_1(\lambda _1, \minus\lambda _2,\minus\omega ),
\\[1mm]
A_8=[I_1(\lambda _1,\lambda _2, \omega )+I_1(\minus\lambda _1,\minus\lambda _2, \omega )],
&
B_8=-I_1(\minus\lambda _1,\lambda _2, \minus\omega ),
&
C_8=-I_1(\lambda _1,\minus\lambda _2, \minus\omega).
\end{array}
\ed
All these formulae may be established by elementary methods. For example,
\bd
\bra x~\sgn (\lambda _2\plus y)e^{-i\hat {S}\bk \cdot \ba }\ket
=\int\! dxdy~x~p(x,y)
\sgn(\lambda _2\plus y)e^{-\frac{1}{2}i\hat{S}\eta a^2[\sgn(\lambda _2+y)-\sgn (\lambda _1 +x)]}
\ed
\bd
=\biggl [\int _{-\infty }^{-\lambda _1}\!dx~x+\int _{-\lambda _1}^\infty\!dx~ x\biggr ]
\biggl [\int _{-\lambda _2}^\infty\!dy~ e^{-\frac{1}{2}i\hat {S}\eta a^2(1-\sgn (\lambda _1 +x))}p(x,y)\biggr ]
~~~~~~~~~~~~
\ed
\bd
~~~~~~~~~~~~
-\biggl [\int _{-\infty }^{-\lambda _1}\!dx~x+\int _{-\lambda _1}^\infty\!dx~ x\biggr ]
\biggl [\int _{\lambda _2}^\infty\!dy~ e^{\frac{1}{2}i\hat {S}\eta a^2(1+\sgn (\lambda _1 +x))}p(x,\minus y)\biggr ]
\ed
\bd
=\int _{\lambda _1 }^\infty\!dx~ x\int _{\lambda _2}^\infty\!dy~ p(x,y)-\int _{-\lambda _1 }^\infty\!dx~ x
\int _{\lambda _2}^\infty\!dy~ e^{i\hat {S}\eta a^2}p(x,\minus y)
~~~~~~~~~~~~
\ed
\bd
~~~~~~~~~~~~
-\int _{\lambda _1 }^\infty\!dx~x\int _{-\lambda _2}^\infty\!dy~e^{-i\hat {S}\eta a^2}p(x,-y)+
\int _{-\lambda _1 }^\infty\!dx~x\int _{-\lambda _2}^\infty\!dy~ p(x,y)
\ed
$$
=I_2(\lambda _1,\lambda _2, \omega )-e^{i\hat {S}\eta a^2}I_2(-\lambda _1,\lambda _2,-\omega )
-e^{-i\hat {S}\eta a^2}I_2(\lambda _1,-\lambda _2,-\omega )+I_2(-\lambda _1,-\lambda _2, \omega )
$$
$$
=A_3+B_3e^{i\hat {S}\eta a^2}+C_3e^{-i\hat {S}\eta a^2},
$$
\no
as required.

\section{Analysis of Macroscopic Distribution in Phase III}
\label{app:phaseIIIdetails}

Here we give the details of our analysis of the macroscopic
distribution $P_\tau(S,Q,R)$ in phase III, starting from
equation (\ref{eq:WPhaseIIandIII3}).
We note that, in phase III:
\bd
e^{-i\hat {S}\beps\cdot \ba }=e^{-i\hat {S}\bk \cdot \ba }\biggl \{
1-\frac{i\eta a z \hat {S}}{2\sqrt {N}}[\sgn(\lambda _2\plus y)\minus \sgn (\lambda _1 \plus x)]
-\frac{(\eta a z \hat {S})^2}{4 N}[1\minus \sgn(\lambda _1\plus x)\sgn (\lambda _2 \plus y)]+\cdots \biggr \}
\ed
The terms which we neglected are  $\order(N^{-2})$, since when performing averages
over the training set the average of the $z^3$ term is zero.
Equation (\ref{eq:WPhaseIIandIII3}) now yields
\bd
{\cal W }_\tau[\bOmega ,\bOmega^\prime]=N\!
\int\!\! \frac{d\hat {\bOmega }}{(2\pi )^3}\bigbra
e^{i\hat {\bOmega }\cdot [\bOmega -\bOmega (\bJ )]}
\bra e^{-i\hat {S}\bk \cdot \ba }\minus 1
\minus \frac{(\eta a \sigma \hat {S})^2}{4N}
e^{-i\hat {S}\bk \cdot \ba }[1\minus \sgn (\lambda _1 \plus x)\sgn (\lambda _2 \plus
y)]\ket_\bxi
\right.
\ed
\be
\left.
~~~~~~~~~~~~~~~~~~~~~
-\int\! \frac{d\hat {\bOmega }}{(2\pi)^3}~
e^{i\hat {\bOmega }\cdot [\bOmega -\bOmega (\bJ )]}\biggl\{\cdots \biggr \}_{\III}
\bigket_{\bOmega^\prime}
\label{WPhaseIII3}
\ee
\no
where
\bd
\biggl\{
\cdots \biggr \}_{\III}=iN\sum _{i\mu }\bra k_i
\frac {\partial \Phi_\mu}{\partial J_i}e^{-i\hat {S}\bk \cdot \ba }
\ket_\bxi {\hat {\Phi }}_\mu +\frac {1}{2} iN\sum _{ij\mu }
\bra  k_i k_j\frac {\partial^2\Phi_\mu}{\partial J_i\partial J_j}
e^{-i\hat{S}\bk \cdot \ba }\ket_\bxi {\hat{\Phi}}_\mu
\ed
\be
+\frac {1}{2}N\sum _{ij\mu\nu }\bra k _i k _j
\frac {\partial \Phi_\mu }{\partial J_i}\frac {\partial \Phi_\nu }{\partial J_j}
e^{-i\hat{S}\bk \cdot \ba }
\ket_\bxi {\hat {\Phi }}_\mu {\hat {\Phi }}_\nu
\label{WPhaseIIandIII4}
\ee
We showed in appendix \ref{app:integrals} that
\bd
\int\! dxdy~p(x,y)\sgn(\lambda _1\plus x)\sgn (\lambda _2\plus y)e^{-i\hat {S}\bk \cdot \ba }
=I_1(\lambda_1,\lambda_2,
\omega)-I_1(\minus \lambda_1, \lambda_2,\minus \omega )e^{i\hat {S}
\eta a^2}
\ed
\bd
-I_1(\lambda_1, \minus \lambda_2, \minus \omega )e^{-i\hat {S}\eta a^2}
+I_1(\minus \lambda_1,\minus \lambda _2,\omega )
\ed
so that
\bd
\int\! dxdy~p(x,y) e^{-i\hat {S}\bk \cdot \ba }[1\minus \sgn (\lambda _1\plus x)
\sgn (\lambda _2 \plus y)]
=2[e^{i\hat {S}\eta a^2}
I_1(\minus \lambda_1, \lambda _2, \minus\omega )+e^{-i\hat {S}\eta a^2}
I_1(\lambda_1, \minus\lambda _2, \minus\omega )],
\ed
by virtue of equation (\ref{eminusishatkdota}) and the fact that $I_1(\lambda _1,\lambda _2,\omega )
+I_1(\minus \lambda_1,\minus\lambda_2 ,\omega )=1\minus E_g$.
Bearing in mind the sub-shell average we may write
\bd
\int\! d\hat{S}~\hat {S}^2~
e^{i\hat {S}[S\pm \eta a^2-S(\bJ )]}=-2\pi \frac{\partial^2}{\partial {S^\prime}^2}
\delta [S\pm \eta a^2\minus S^\prime]
\ed
Upon combining equations (\ref{eq:newdPOmegadt},\ref{eminusishatkdota},\ref{WPhaseIII3}) we
find that in phase III the joint probability density $P_\tau (S,Q,R)$ satisfies
\bd
\frac {d}{d\tau}P_\tau(S,Q,R)=N\left[\room
I_1(\minus \lambda_1(S^+),\lambda_2 ,\minus \omega )P_\tau(S^+,Q,R)
+I_1(\lambda_1(S^-),\minus \lambda_2 ,\minus\omega
)P_\tau(S^-,Q,R)
\right.
\ed
\bd
\left.
-E_g(S,Q,R)P_\tau(S,Q,R)\room\right]
+\frac {1}{2}\eta ^2a^2\sigma^2 \int\! d\bOmega^\prime~
P_\tau (\bOmega^\prime)\left[\room
I_1(\minus \lambda_1(S^\prime),\lambda_2, \minus\omega )\frac{\partial^2}{\partial {S^\prime}^2}
\delta [S\plus \eta a^2\minus S^\prime]\right.
\ed
\bd
\left.
+I_1(\lambda _1(S^\prime), \minus\lambda_2, \minus\omega )\frac{\partial^2}{\partial {S^\prime}^2}
\delta [S\minus \eta a^2\minus S^\prime]\right]
\delta [\bPhi \minus \bPhi^\prime]
-\int\!\frac{ d\bOmega^\prime}{(2\pi)^3}~ P_\tau (\bOmega^\prime)\bigbra \int\! d\hat {\bOmega
}~e^{i\hat {\bOmega }\cdot [\bOmega -\bOmega (\bJ )]}\biggl\{\cdots \biggr \}_{\III}\bigket
_{\bOmega^\prime}
\ed
where $\biggl\{\cdots \biggr \}_{\III}$ is given by equation
(\ref{WPhaseIIandIII4}), and hence
\bd
\frac{d}{d\tau}P_\tau (S,Q,R)=N\left[\room
I_1(\minus\lambda_1(S^+),\lambda_2 ,\minus\omega ) P_\tau(S^+,Q,R)
+I_1(\lambda_1(S^-),\minus \lambda_2 ,\minus \omega) P_\tau(S^-,Q,R)
\right.
\ed
\bd
\left.
-E_g( {S},Q,R)P_\tau(S,Q,R)\room\right]
+\frac {1}{2}\eta ^2a^2\sigma ^2\biggl[
\frac{\partial^2}{\partial S^2} [I_1(\minus\lambda _1(S^+), \lambda_2, \minus\omega) P_\tau (S^+,Q,R)]
\ed
\be
+\frac{\partial^2}{\partial S^2}[I_1(\lambda_1(S^-),\minus\lambda_2, \minus\omega) P_\tau (S^-,Q,R)]
\biggr ]
-\int\! \frac{d\bOmega^\prime}{(2\pi)^3}~P_\tau (\bOmega
') \int\! d\hat{\bOmega }\bigbra
e^{i\hat {\bOmega }\cdot [\bOmega -\bOmega (\bJ )]}\biggl\{\cdots \biggr \}_{\III}\bigket
_{\bOmega^\prime}
\label{dPdtau10}
\ee
As regards the evaluation of $\biggl\{\cdots \biggr \}_{\III}$ we note that
\bd
N\sum _{i\mu }\bra k_i
\frac {\partial \Phi_\mu}{\partial J_i}e^{-i\hat {S}\bk \cdot \ba }
\ket_\bxi {\hat {\Phi }}_\mu=\eta J\bra (\lambda _1\plus x)[\sgn (\lambda_2
\plus y)\minus \sgn (\lambda _1\plus x)]e^{-i\hat {S}\bk \cdot \ba }\ket{\hat {\Phi }}_1
\vspace*{-2mm}
\ed
\bd
+\frac{1}{2}\eta \bra (\lambda _2\plus x)[\sgn (\lambda_2
\plus y)\minus \sgn (\lambda _1\plus x)]e^{-i\hat {s}\bk \cdot \ba }\ket {\hat {\Phi }}_2
\ed
\bd
=\eta J[K_1\plus L_1e^{i\hat {s}\eta a^2}\plus M_1e^{-i\hat {s}\eta a^2}]{\hat {\Phi }}_1
+\frac {1}{2}\eta [K_2\plus L_2e^{i\hat {s}\eta a^2}\plus M_2e^{-i\hat {S}\eta a^2}]{\hat {\Phi }}_2
\ed
in which
\bd
K_1=\lambda _1(A_1\minus A_2)\plus (A_3\minus A_4),~~~~~~K_2=\lambda _2(A_1\minus A_2)\plus (A_5\minus A_6),
\ed
\bd
L_1=\lambda _1(B_1\minus B_2)\plus (B_3\minus B_4),~~~~~~L_2=\lambda _2(B_1\minus B_2)\plus (B_5\minus B_6),
\ed
\bd
M_1=\lambda _1(C_1\minus C_2)\plus (C_3\minus C_4),~~~~~~M_2=\lambda _2(C_1\minus C_2)\plus (C_5\minus C_6)
\ed
and $A_i,~B_i,~C_i$ are functions defined in appendix \ref{app:integrals} and expressed in terms of the integrals
$I_1(\lambda_1,\lambda _2 ,\omega )$ and $I_2(\lambda _1,\lambda _2 ,\omega
)$.
In a similar way we find that
\bd
N\sum _{ij\mu }\bra k_i k_j\frac {\partial^2\Phi _\mu }{\partial J_i\partial J_j}
e^{-i\hat {S}\bk \cdot \ba }\ket_\bxi {\hat {\Phi }}_\mu=\eta ^2\bra [1\minus\sgn (\lambda _1\plus x)
\sgn (\lambda _2\plus y)]
e^{i\hat {s}\bk \cdot \ba }\ket {\hat {\Phi }}_1
\vspace*{-2mm}
\ed
\bd
=\eta ^2[K_3+L_3e^{i\hat {s}\eta a^2}+M_3^{-i\hat {s}\eta a^2}]{\hat {\Phi }}_1
\ed
where $K_3=A_7\minus A_8$, $L_3=B_7\minus B_8$, and $M_3=C_7\minus
C_8$.
Note that the term
\bd
N\sum _{ij\mu\nu }\bra k_i k _j
\frac {\partial \Phi_\mu }{\partial J_i}\frac {\partial \Phi_\nu }{\partial J_j}e^{-i\hat {S}\bk \cdot \ba }
\ket_\bxi {\hat {\Phi }}_\mu {\hat {\Phi }}_\nu
\ed
\no
makes no contribution to ${\cal W}_\tau[\bOmega ,\bOmega^\prime]$ in the
limit of large $N$. Using equations (\ref{WPhaseIIandIII4}) and (\ref{dPdtau10}) we can now carry
out the remaining integrations using standard formulae from distribution theory, as described for earlier
phases,
and find that
\bd
\frac {d}{d\tau}
P_\tau(S,Q,R)=\hspace*{130mm}
\ed
\bd
N\left[\room
I_1(\minus\lambda _1(S^+),\lambda _2 ,\minus\omega ) P_\tau (S^+\!,Q,R)
+I_1(\lambda _1(S^-),-\lambda _2 ,-\omega ) P_\tau(S^-\!,Q,R)
-E_g(S,Q,R)P_\tau(S,Q,R)\right]
\ed
\bd
+\frac {1}{2}\eta ^2a^2\sigma ^2\biggl [\frac{\partial^2}{\partial S^2}
[I_1(\minus\lambda_1(S^+), \lambda_2, \minus\omega )P_\tau (S^+\!,Q,R)]
+\frac{\partial^2}{\partial S^2}[I_1(\lambda _1(S^-), \minus\lambda_2, \minus\omega)P_\tau(S^-\!,Q,R)]
\biggr ]
\ed
\bd
-\frac {\partial }{\partial Q}\left[\room
\eta J[K_1P_\tau ({S},Q,R)\plus L_1 P_\tau (S^+\!,Q,R)\plus M_1
P_\tau(S^-\!,Q,R)]
\right.
\ed
\bd
\left.
+\frac {1}{2}\eta ^2[K_3P_\tau ({S},Q,R)\plus L_3P_\tau (S^+\!,Q,R)\plus M_3 P_\tau
(S^-\!,Q,R)]
\room\right]
\ed
\be
-\frac {\partial}{\partial R}\biggl [
\frac {1}{2}\eta [K_2P_\tau ({S},Q,R)\plus L_2P_\tau (S^+,Q,R)\plus M_2P_\tau (S^-,Q,R)]
\biggr ]
\label{PDEforPtauSQR}
\ee
Integration over $S$ now gives, in combination with the relation
$E_g=I_1(\minus \lambda_1, \lambda_2, \minus \omega )+I_1(\lambda_1, \minus\lambda _2,
\minus\omega )$:
\bd
\frac {d}{d\tau} P_\tau(Q,R)=
-\frac {\partial}{\partial Q}\left\{
\room
P_\tau (Q,R)\left[
\eta J\int\! dS~(K_1\plus L_1\plus M_1) P_\tau ({S}|Q,R)
~~~~~~~~~~~~~~~~~~~~~~~~~~~~~~~~~~~~~~~~~~~~~~~~~~~~~~~
\right.\right.
\ed
\bd
\left.\left.
+\frac {1}{2}\eta ^2\int\! dS~ (K_3\plus L_3\plus M_3)P_\tau(S|Q,R)\right]
\right\}
-\frac {\partial }{\partial R}\left\{
\room
P_\tau(Q,R)\left[
\frac {1}{2}\eta \int\! dS ~(K_2\plus L_2\plus
M_2)P_\tau(S|Q,R)\right]\right\}
\ed
which is a Liouville equation with solution $P_\tau (Q,R)=\delta [Q\minus Q(\tau)]\delta[R\minus R(\tau)]$,
where the deterministic flow trajectories $(Q(\tau),R(\tau))$ are given
as the solutions of
(\ref{QdotPhaseIII},\ref{RdotPhaseIII}), as claimed.

\end{document}